\documentclass[aps,prl, showpacs,twocolumn]{revtex4}
\usepackage{amssymb}
\usepackage{amsmath}
\usepackage{graphicx}
\usepackage{epsfig}

\begin{document}

\title{Dynamic generation of entangling wave packets \\
in $XY$ spin system with decaying long range couplings}
\author{S. Yang$^{1}$, Z. Song$^{1,a}$ and C.P. Sun$^{1,2,a,b}$ }
\affiliation{$^{1}$Department of Physics, Nankai University, Tianjin 300071, China}
\affiliation{$^{2}$ Institute of Theoretical Physics, Chinese Academy of Sciences,
Beijing, 100080, China}

\begin{abstract}
We study the dynamic generation of spin entanglement between two distant
sites in an $XY$ model with $1/r^{2}$-decay long range couplings. Due to the
linear dispersion relation $\varepsilon (k)\sim |k|$ of magnons in such
model, we show that a well-located spin state can be dynamically split into
two moving entangled local wave packets without changing their shapes.
Interestingly, when such two wave packets meet at the diametrically opposite
site after the fast period $\tau =\pi $, the initial well-located state can
be recurrent completely. Numerical calculation is performed to confirm the
analytical result even the ring system of sizes $N$ up to several thousands
are considered. The truncation approximation for the coupling strengths is
also studied. Numerical simulation shows that the above conclusions still
hold even the range of the coupling strength is truncated at a relative
shorter scale comparing to the size of the spin system.
\end{abstract}

\pacs{03.67.-a, 03.67.Lx, 03.65.Ud, 75.10.Jm}
\maketitle

\section{Introduction}

In quantum information processing, it is also\ crucial to generate entangled
qubits, which can be used to perfectly transfer a quantum state over long
distance. For optical system this task has been completed long time ago, but
for a solid state systems it remains a great challenge both in experiment
and theoretical setup to create quantum entanglement by a solid state
device. Recently many proposals to entangle distant spins have been proposed
based on various physical mechanisms \cite{Bose1,Paternostro,YS1,CB}. Most
protocols for accomplishing quantum state transfer in a spin array base on
the fixed inter-qubit couplings \cite%
{Subra,Bose2,Ekert,TJ,key-17,key-18,BBG,key-6,Plenio,ST,LY,YS2,qian,LY2}.
The simplest coupled spin system with uniform nearest neighbor (NN) coupling
has been studied by the pioneer work \cite{Bose2}. The paradigm\ to generate
maximal entanglement and to perform perfect qubit-state transmission over
arbitrary distance is the protocols using the pre-engineered inhomogeneous
NN couplings \cite{Ekert,ST}. Among them, the practical scheme realizes the
transmission of Gaussian wave packet as flying qubit via the spin system
with uniform NN couplings \cite{TJ,YS2}. The advantages of such scheme rest
with its fast transfer, which means that the period of transfer time is
proportional to the distance, and the higher fidelity for longer distance.
We also notice that there is a current interest in studying systems with
long-range inter-qubit interactions \cite{BeyondNN,Bose3,longrange}.

In this paper, we revisit the issue of entanglement generation and quantum
state transfer between two distant qubits in a qubit array with long-range
inter-qubit interactions. An alternative way to construct a perfect medium
for quantum state transfer may require the long-range interactions beyond NN
interactions, but they should be required to decay rapidly in order to avoid
the direct connecting coupling between two distant qubits that trivially
causes quantum entanglement. Now we propose a novel protocol based on a
pre-engineered $XY$-model with long-range $1/r^{2}$-decay interactions. It
is found that such model has the same function as that of the modulated NN
coupling spin model \cite{Ekert}, offering significant advantages over other
protocols in the tasks perfectly transferring quantum state and generating
entanglement between two sites over longer distance.

This paper is organized as follows, in Sec. II, we consider a model with
linear dispersion relation and show that it ensures the perfect state
transfer and creation of entanglement between spatial separated qubits. In
Sec. III we propose a spin model with $1/r^{2}$-decay interactions that
process the desired dispersion relation. In Sec. IV we perform numerical
calculation to confirm some of the obtained analytical results and the
validity of the truncation approximation.
\begin{figure}[tbp]
\includegraphics[bb=108 319 505 600, width=7 cm, clip]{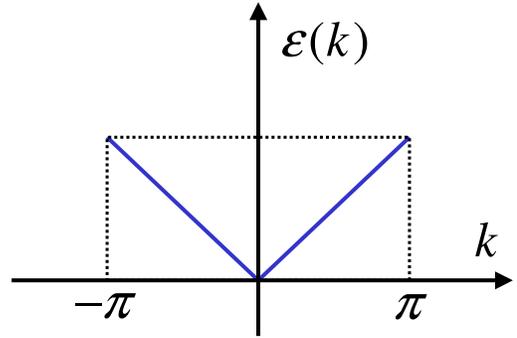}
\caption{\textit{(color online) Schematic illustration for the ideal
dispersion relation $\protect\varepsilon (k)\sim |k|$ of a system which is
shown to be a perfect entangler. Such kind of system can be realized by the $%
XY$ model with pre-engineered long-range couplings.}}
\end{figure}
\begin{figure}[tbp]
\includegraphics[bb=75 346 555 723, width=8 cm, clip]{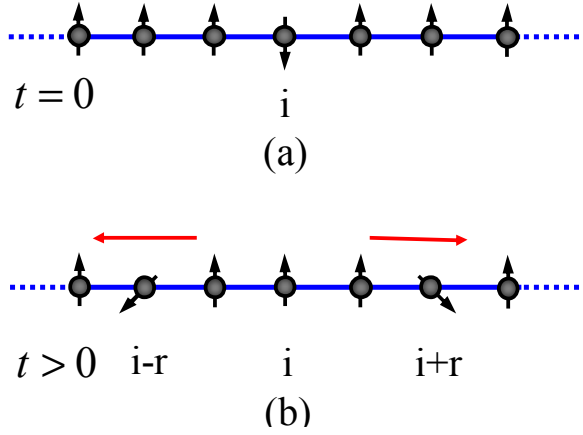}
\caption{\textit{(color online) Schematic illustration of the generation
process of two distant entangled qubits. At $t=0$ the initial state is a
single-spin flip on the saturated ferromagnet. At $t>0$ the initial $\protect%
\delta $-pulse magnon is separated into two components which have maximal
entanglement. At $t=\protect\tau $, the recurrence time, the two sub-wave
packets meet together and the initial state recurrent.}}
\end{figure}

\section{Formalism}

\subsection{Pre-engineered model with linear dispersion relation}

Usually, we regard a $1/2$ spin as a qubit, and a coupled spin system as a
qubit array respectively. The simple coupled qubit array is usually
described by the spin-$1/2$ $XY$ model. Our proposal makes the quantum spin
array behave as a perfect spin-networks, in which a spin flip at any site
(or the superposition of local single-spin-flip states) can evolve into two
entangled, local wave packets. Consider the prescribed spin-$1/2$ $XY$-model
on ring system with $N$ sites. The Hamiltonian reads

\begin{equation}
H=2\sum_{i,r}J_{r}(S_{i}^{x}S_{i+r}^{x}+S_{i}^{y}S_{i+r}^{y})  \label{H1}
\end{equation}%
or%
\begin{equation}
H=\sum_{i,r}J_{r}(S_{i}^{+}S_{i+r}^{-}+S_{i}^{-}S_{i+r}^{+}),
\end{equation}%
where $S_{i}^{x}$, $S_{i}^{y}$ and $S_{i}^{z}$ are Pauli matrices for the
site $i$th, and $S_{i}^{\pm }=S_{i}^{x}\pm iS_{i}^{y}$; $J_{r}$ is the
coupling strength for the two spins separated by the distance $r$. Since the
Hamiltonian conserves spin, i.e., $[\sum_{i}S_{i}^{z},H]=0$, the dynamics
can be reduced to that in some invariant subspaces. Thus we can only
concentrate on the single excitation subspace hereafter. If a single-site
flipped state can be correctly transferred, a qubit state should also be
transferred correspondingly because the saturate ferromagnetic state with
all spins up state $\left\vert 0\right\rangle \equiv
\prod_{i=1}^{N}\left\vert \uparrow \right\rangle _{i}$ is an eigenstate of
the Hamiltonian. Actually, in the single-spin-flip subspace or the single
excitation (magnon) subspace, the basis states are denoted by the
single-site flipped state$\ $(or $\delta $-pulse) $\left\vert i\right\rangle
=S_{i}^{-}\left\vert 0\right\rangle $, $i\in \lbrack 1,N]$. Therefore, if
state $\left\vert i\right\rangle $\ can evolve to state $\left\vert
j\right\rangle $\ after a period of time $\tau $,\ $\left\vert
j\right\rangle =\exp (-iH\tau )\left\vert i\right\rangle $,\ we have

\begin{equation}
(\alpha \left\vert \uparrow \right\rangle _{j}+\beta \left\vert \downarrow
\right\rangle _{j})\prod_{l\neq j}^{N}\left\vert \uparrow \right\rangle
_{l}=e^{-iH\tau }(\alpha \left\vert \uparrow \right\rangle _{i}+\beta
\left\vert \downarrow \right\rangle _{i})\prod_{l\neq i}^{N}\left\vert
\uparrow \right\rangle _{l},
\end{equation}%
i.e., a qubit state $\alpha \left\vert \uparrow \right\rangle +\beta
\left\vert \downarrow \right\rangle $ can be transferred from $i$ to $j$.
Furthermore, any single-magnon state can be expressed as

\begin{equation}
\left\vert \psi \right\rangle =\sum\limits_{i}A_{i}S_{i}^{-}\left\vert
0\right\rangle \equiv \sum\limits_{i}A_{i}\left\vert i\right\rangle
\end{equation}%
and
\begin{equation}
\left\vert \psi \right\rangle =\sum\limits_{k}D_{k}S_{k}^{-}\left\vert
0\right\rangle \equiv \sum\limits_{k}D_{k}\left\vert k\right\rangle
\end{equation}%
respectively in spatial $\{\left\vert i\right\rangle \}$\ and momentum $%
\{\left\vert k\right\rangle \}$ spaces. Here, we have used the spin-wave
operator%
\begin{equation}
S_{k}^{-}=\frac{1}{\sqrt{N}}\sum_{j}e^{-ikj}S_{j}^{-},
\end{equation}%
with discrete momentum $k=2\pi n/N$, $n\in \lbrack -N/2,N/2-1]$.

We begin with assumption that there exists an optimal distribution of $J_{r}$%
, which ensures the single-magnon spectrum possessing a linear dispersion
relation, i.e.%
\begin{eqnarray}
H_{s} &=&\sum\limits_{k}\varepsilon _{k}\left\vert k\right\rangle
\left\langle k\right\vert  \label{dispersion} \\
\varepsilon _{k} &=&\frac{N}{2\pi }\left\vert k\right\vert  \notag
\end{eqnarray}%
as illustrated in Fig. 1. In the following, we first show that such kind of
systems can perform perfect state transfer and long range entanglement
generation, and then we can provide a practical example that satisfies this
linear dispersion relation.

\subsection{Time evolution of wave packets in the linear dispersion regime}

In order to investigate the dynamics of\ generating entanglement in the spin
system with linear dispersion relation mentioned above, we concentrate on
the case where the initial state is single-site flipped state$\
S_{i}^{-}\left\vert 0\right\rangle $, $i\in \lbrack 1,N]$. Intuitively, a
well localized state has equal probability amplitudes with respect to
momentum eigenstates $|\pm \left\vert k\right\vert \rangle $, due to the
inverse Fourier transformation

\begin{eqnarray}
S_{j}^{-} &=&\frac{1}{\sqrt{N}}\sum_{k}e^{ikj}S_{k}^{-}  \notag \\
&&\underrightarrow{N\longrightarrow \infty }\frac{1}{\sqrt{N}}\frac{N}{2\pi }%
\int_{-\pi }^{\pi }e^{ikj}S_{k}^{-}dk  \notag \\
&=&\frac{\sqrt{N}}{2\pi }\int_{0}^{\pi
}(e^{ikj}S_{k}^{-}+e^{-ikj}S_{-k}^{-})dk.
\end{eqnarray}

Thus it should be split into two classes of waves with $\pm \left\vert
k\right\vert $ driven by the Hamiltonian with linear single-magnon
dispersion relation (\ref{dispersion}). If the superposition of the two
classes of waves are still well-localized (or form two wave packets) in real
space, it will lead to the entanglement of the two wave packets.

Generally we first consider a wave packet dynamics with an initial state%
\begin{equation}
\left\vert \psi (N_{A},0)\right\rangle =S_{N_{A}}^{-}\left\vert
0\right\rangle =\frac{1}{\sqrt{N}}\sum_{k}e^{ikN_{A}}S_{k}^{-}\left\vert
0\right\rangle ,  \label{initial}
\end{equation}%
which is a single flip at site $N_{A}$. The time evolution starting with
this state can be calculated as
\begin{eqnarray}
\left\vert \psi (N_{A},t)\right\rangle  &=&e^{-iHt}\left\vert \psi
(N_{A},0)\right\rangle   \notag \\
&=&\left\vert \phi _{-}(N_{A},t)\right\rangle +\left\vert \phi
_{+}(N_{A},t)\right\rangle
\end{eqnarray}%
where%
\begin{equation}
\left\vert \phi _{\pm }(N_{A},t)\right\rangle =\mp \frac{\sqrt{N}}{2\pi }%
\int_{0}^{\mp \pi }e^{ik(N_{A}\pm \frac{N}{2\pi }t)}dk\left\vert
k\right\rangle
\end{equation}%
denote the two wave packets with explicit forms in the real space%
\begin{eqnarray}
\left\vert \phi _{\pm }(N_{A},t)\right\rangle  &=&\mp \frac{\sqrt{N}}{2\pi }%
\int_{0}^{\mp \pi }e^{ik(N_{A}\pm \frac{N}{2\pi }t)}dk\frac{1}{\sqrt{N}}%
\sum_{j}e^{-ikj}\left\vert j\right\rangle   \notag \\
&=&\frac{1}{2\pi }\sum_{j}\left[ \frac{\mp e^{\mp i\pi (N_{A}-j\pm \frac{N}{%
2\pi }t)}\pm 1}{i\left( N_{A}-j\pm \frac{N}{2\pi }t\right) }\right]
\left\vert j\right\rangle   \label{subwaves}
\end{eqnarray}

We will show that such two wave packets are localized and nonspreading, with
velocity $v_{\pm }=\pm N/2\pi $,\ respectively. Actually, from Eq. (\ref%
{subwaves}) one have

\begin{equation}
\left\vert \phi _{\pm }(N_{A},t+t_{0})\right\rangle =\left\vert \phi _{\pm
}(N_{A}\pm \frac{N}{2\pi }t,t_{0})\right\rangle
\end{equation}%
or
\begin{equation}
e^{-iHt}\left\vert \phi _{\pm }(N_{A},t_{0})\right\rangle =T_{\frac{N}{2\pi }%
t}\left\vert \phi _{\pm }(N_{A},t_{0})\right\rangle  \label{tran}
\end{equation}%
where $T_{a}$ acts as translational operator for the states $\left\vert \phi
_{\pm }(l,t)\right\rangle $: $T_{a}\left\vert \phi _{\pm }(l,t)\right\rangle
\equiv \left\vert \phi _{\pm }(l\pm a,t)\right\rangle $. In general, the
spatial coordinate is treated as discrete variable while the temporal
coordinate is continuous. In order to make $T_{a}$\ to be operable, the
translational spacing $a$\ should be integer. In the following, we only
consider the states at discrete instant $t$, which ensures $a$\ to be
integer, i.e., $Nt/2\pi =\left[ Nt/2\pi \right] $ (the integer part of $%
Nt/2\pi $). In terms of
\begin{equation}
l_{\pm }(t)=N_{A}-j\pm \left[ \frac{N}{2\pi }t\right] ,
\end{equation}%
the wave packets (\ref{subwaves}) can be re-written as
\begin{equation}
\left\vert \phi _{\pm }(N_{A},t)\right\rangle =\frac{1}{2\pi }\sum_{j}\{%
\frac{1}{il_{\pm }(t)}[\mp e^{\mp i\pi l_{\pm }(t)}\pm 1]\}\left\vert
j\right\rangle .
\end{equation}

Obviously, Eq. (\ref{tran}) shows that state $\left\vert \phi _{\pm
}(N_{A},t)\right\rangle $ is shape-invariant with velocity $v_{\pm }=\pm
N/2\pi $. Now we show the locality of these states. The projections
\begin{equation}
\left\langle j\right. \left\vert \phi _{\pm }(N_{A},t)\right\rangle =\left\{
\begin{array}{c}
\frac{1}{2},\text{ \ \ \ \ }l_{\pm }(t)=0 \\
\mp \frac{i}{\pi l_{\pm }(t)},\text{ \ odd }l_{\pm }(j,t) \\
0,\text{ \ \ even }l_{\pm }(t)\neq 0%
\end{array}%
\right.
\end{equation}%
of $\left\vert \phi _{\pm }(N_{A},t)\right\rangle $ onto the localized state
$\left\vert j\right\rangle $ shows that the shape-invariant states are
well-localized. And each wave packet has the probability amplitude

\begin{equation}
\left\vert \phi _{\pm }(N_{A},t)\right\vert ^{2}\simeq \frac{1}{4}+\frac{2}{%
\pi ^{2}}\sum_{n=0}^{\infty }\frac{1}{(2n+1)^{2}}=\frac{1}{2},
\end{equation}%
which indicates that the initial state (\ref{initial})\ is splited into two
wave packets completely. On the other hand, the dispersion (\ref{dispersion}%
) requires the translational invariance of the Hamiltonian since the
momentum is the conserved quantity, which leads to the periodicity of the
states $\left\vert \phi _{\pm }\right\rangle $

\begin{equation}
\left\vert \phi _{\pm }(N_{A}\pm \frac{N}{2\pi }t,t)\right\rangle
=\left\vert \phi _{\pm }(N_{A}\pm \frac{N}{2\pi }t\mp m_{\pm
}N,t)\right\rangle ,
\end{equation}%
where $m_{\pm }=1,2,\cdots $, are integers. Obviously, when $\tau
=(m_{+}+m_{-})\pi $, the initial state $\left\vert \psi
(N_{A},0)\right\rangle =S_{N_{A}}^{-}\left\vert 0\right\rangle $ evolves to%
\begin{equation}
\left\vert \psi (N_{A},\tau )\right\rangle =S_{N_{A}-\frac{N}{2}%
(m_{+}-m_{-})}^{-}\left\vert 0\right\rangle ,
\end{equation}%
which is just the recurrence of $\left\vert \psi (N_{A},0)\right\rangle $ on
the positions $N_{A}-N(m_{+}-m_{-})/2$ at instants $(m_{+}+m_{-})\pi $. The
physics of this phenomenon can be understood as the interference of two wave
packets (\ref{subwaves}). It also accords with the prediction for the system
with spectrum-symmetry matching condition (SSMC) \cite{ST,LY2}.

In the single magnon invariant subspace, the single flip states $%
\{\left\vert j\right\rangle \}$ at site $j$\ constitute the complete basis.\
Then the above conclusion can be applied to any states in this subspace.
Consider an arbitrary initial state $\left\vert \Phi (0)\right\rangle
=\sum_{j}A_{j}\left\vert j\right\rangle $ $=\sum_{j}A_{j}\left\vert \psi
(j,0)\right\rangle $, which is a coherent superposition of single flip
states. At time $t$, it evolves to%
\begin{equation}
\left\vert \Phi (t)\right\rangle =e^{-iHt}\left\vert \Phi (0)\right\rangle
=\left\vert \Phi _{+}(t)\right\rangle +\left\vert \Phi _{-}(t)\right\rangle ,
\end{equation}%
where%
\begin{equation}
\left\vert \Phi _{\pm }(t)\right\rangle =\sum_{j}A_{j}\left\vert \phi _{\pm
}(j,t)\right\rangle
\end{equation}%
represent two invariant-shape states. Accordingly, at instants $\tau
=(m_{+}+m_{-})\pi $, the final state is a translation of $\left\vert \Phi
(0)\right\rangle $

\begin{equation}
\left\vert \Phi (\tau )\right\rangle =T_{\frac{N}{2}(m_{+}-m_{-})}\left\vert
\Phi (0)\right\rangle .
\end{equation}%
Furthermore, during the period of time $t\neq \tau $, wave packets $%
\left\vert \Phi _{\pm }(t)\right\rangle $ are still well localized in space
if the initial state $\left\vert \Phi (0)\right\rangle $ is local. Then the
revival of a wave packet can be used to implement perfect quantum state
transfer.

\subsection{Entanglement of two separated spins}

Now we turn our attention on the entanglement of the two separated spins
induced by the wave packets (\ref{subwaves}). The reduced density matrix of
a state $\left\vert \Phi (t)\right\rangle $ for two spins located at sites $%
i $ and $j$ \cite{Wootters,Wang} has the form
\begin{equation}
\rho _{ij}=\left(
\begin{array}{cccc}
v_{ij}^{+} & 0 & 0 & 0 \\
0 & w_{ij} & z_{ij} & 0 \\
0 & z_{ij} & w_{ij} & 0 \\
0 & 0 & 0 & v_{ij}^{-}%
\end{array}%
\right)
\end{equation}%
with respect to the standard basis vectors $\left\vert \uparrow \uparrow
\right\rangle $, $\left\vert \uparrow \downarrow \right\rangle $, $%
|\downarrow \uparrow \rangle $, and $\left\vert \downarrow \downarrow
\right\rangle $. Here, the matrix elements are%
\begin{eqnarray}
v_{ij}^{\pm } &=&\frac{1}{4}+\frac{1}{2}\left\langle \Phi (t)\right\vert
[2S_{i}^{z}S_{j}^{z}\pm (S_{i}^{z}+S_{j}^{z})]\left\vert \Phi
(t)\right\rangle ,  \notag \\
w_{ij} &=&\frac{1}{4}-\left\langle \Phi (t)\right\vert
S_{i}^{z}S_{j}^{z}\left\vert \Phi (t)\right\rangle ,  \notag \\
z_{ij} &=&\frac{1}{2}\left\langle \Phi (t)\right\vert
(S_{i}^{+}S_{j}^{-}+S_{i}^{-}S_{j}^{+})\left\vert \Phi (t)\right\rangle .
\end{eqnarray}%
Correspondingly, the concurrence of two spins located at sites $i$ and $j$
for a state $\left\vert \Phi (t)\right\rangle $\ can be calculated by
\begin{equation}
C_{ij}=\max \left\{ 0,2\left( \left\vert z_{ij}\right\vert -\sqrt{%
v_{ij}^{+}v_{ij}^{-}}\right) \right\} ,
\end{equation}%
Since the state we concern is in the invariant subspace with $S_{z}=N/2-1$,
we have $v_{ij}^{+}=0$ and then the concurrence reduces to
\begin{equation}
C_{ij}=\left\vert \left\langle \Phi (t)\right\vert
(S_{i}^{+}S_{j}^{-}+S_{i}^{-}S_{j}^{+})\left\vert \Phi (t)\right\rangle
\right\vert .
\end{equation}%
Consider an initial state having reflectional symmetry with respect to a
point $N_{A}$, which has the form

\begin{equation}
\left\vert \Phi (0)\right\rangle =f_{N_{A}}(0)\left\vert N_{A}\right\rangle
+\sum_{j=1}^{N/2}f_{j}(0)[\left\vert N_{A}+j\right\rangle
+(-1)^{R}\left\vert N_{A}-j\right\rangle ],
\end{equation}%
where $R$=even (odd) represents the parity of the state under the
reflection. The concurrence between two spins at $N_{A}+j$ and $N_{A}-j$\
for the state $\left\vert \Phi (t)\right\rangle $ is%
\begin{eqnarray}
C(j,t) &=&C_{N_{A}+j,N_{A}-j}  \notag \\
&=&\left\vert \left\langle \Phi (t)\right\vert
(S_{N_{A}+j}^{+}S_{N_{A}-j}^{-}+S_{N_{A}+j}^{-}S_{N_{A}-j}^{+})\left\vert
\Phi (t)\right\rangle \right\vert  \notag \\
&=&2\left\vert (-1)^{R}f_{j}^{\ast }(t)f_{j}(t)\right\vert  \notag \\
&=&2\left\vert f_{j}(t)\right\vert ^{2}.
\end{eqnarray}%
Obviously, the total concurrence $\sum_{j>0}C(j,t)=1$ due to the normality
of the wave function $\left\vert \Phi (t)\right\rangle $. Usually, the total
concurrence is regarded as the concurrence of two wave packets. Taking the
single flip at $N_{A}$ in the form (\ref{initial}) to be the initial state
as an example, straightforward calculation shows that non-zero concurrences
are%
\begin{eqnarray}
C(j,t_{0}) &=&\frac{1}{2}  \notag \\
C(j,t_{n}) &=&\frac{2}{(2n-1)^{2}\pi ^{2}}.  \label{cj}
\end{eqnarray}%
For this to occur, one requires that $t_{0}=2\pi j/N$, $t_{n}=2\pi \lbrack
j+(2n-1)]/N$, $n=1,2,...$. It indicates that the concurrence with magnitudes
(\ref{cj}) for two spins at $N_{A}+j$ and $N_{A}-j$ can be generated at the
moment $t_{0}$, $t_{1}$, $t_{2}$, ..., $t_{n}$.

\section{Microscopic long-range interaction model}

In this section, we consider the possibility to realize the above scheme in
a system which possesses the linear dispersion relation based on an
one-dimensional arrangement of spins (qubits) coupled by long-range
interactions. Actually, using the identity

\begin{equation}
\left\vert 2\xi \right\vert =\frac{N}{2}-\sum_{r=odd}^{\infty }\frac{4N}{%
r^{2}\pi ^{2}}\cos \frac{r\pi (2\xi )}{N},
\end{equation}%
we have

\begin{equation}
\frac{N}{2\pi }\left\vert k\right\vert \approx
J_{0}+\sum_{r=odd}^{N/2-1}2J_{r}\cos (kr),
\end{equation}%
where

\begin{equation}
J_{0}=\frac{N}{4},\text{ }J_{r}=\frac{N}{r^{2}\pi ^{2}}.  \label{LR}
\end{equation}

Then the Hamiltonian matches the case (\ref{dispersion}) in single-magnon
invariant subspace\ can be rewritten as%
\begin{eqnarray}
H &=&h_{0}+\sum_{r=odd}^{N/2-1}J_{r}h_{r},  \notag \\
h_{0} &=&NJ_{0},  \notag \\
h_{r} &=&\sum_{i}(S_{i}^{+}S_{i+r}^{-}+h.c.).
\end{eqnarray}%
Notice that, although such model involves the long range interaction, $J_{r}$
decays rapidly as $r$ increases. So we call this as \textquotedblleft
microscopic long-range interactions\textquotedblright . The above analytical
analysis shows that such microscopic long-range interaction can lead to
nontrivial long-range entanglement and QST.

\section{Numerical results and truncation approximation}

In\ the previous sections, it is found that the microscopic long-range
interaction can lead to nontrivial long-range entanglement and QST for large
$N$ limit system. In this section, numerical simulations are performed for
finite systems to illustrate the result we obtained above and\ investigate
the application of it. The numerical exact diagonalization method is
employed to calculate the time evolution of the single flip state (\ref%
{initial}) and the Gaussian wavepacket in the following form

\begin{equation}
\left\vert \psi _{G}(N_{A},\alpha )\right\rangle =\frac{1}{\sqrt{\Omega }}%
\sum\limits_{i}e^{-\frac{\alpha ^{2}}{2}(N_{A}-i)^{2}}S_{i}^{-}\left\vert
0\right\rangle  \label{GWP}
\end{equation}%
in the finite $N$ system. Here, $\Omega $\ is the normalized factor and $%
N_{A}$, $\alpha $ determine the center and the shape of wavepacket. In Fig.
3 and 4, the time evolution of a single flip state (\ref{initial}) and a
Gaussian wave packet (\ref{GWP}) of $\alpha =0.1$ in a $N=100$ ring system
are plotted. It shows that the local initial states split into two wave
packets in both cases and the wave packets keep their shapes without
spreading. Similarly, numerical simulations are also employed for the open
chain systems. The similar conclusions are also obtained for open chain
system which will be discussed in the following.

\begin{figure}[tbp]
\includegraphics[bb=30 300 580 670, width=8 cm, clip]{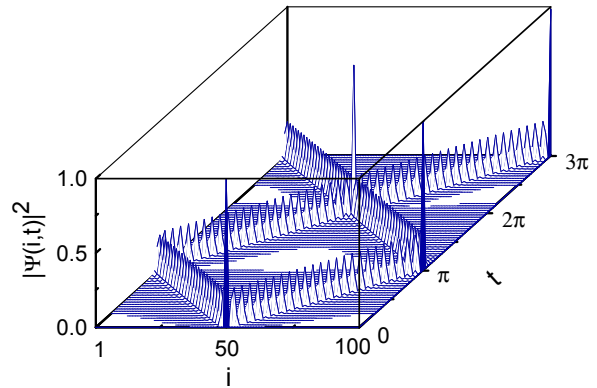}
\caption{\textit{(color online) Time evolution of a $\protect\delta $-pulse
in a $N=100$ ring system obtained by numerical simulation. It shows that the
two sub-waves are local and keep the shape.}}
\end{figure}
\begin{figure}[tbp]
\includegraphics[bb=25 300 580 670, width=8 cm, clip]{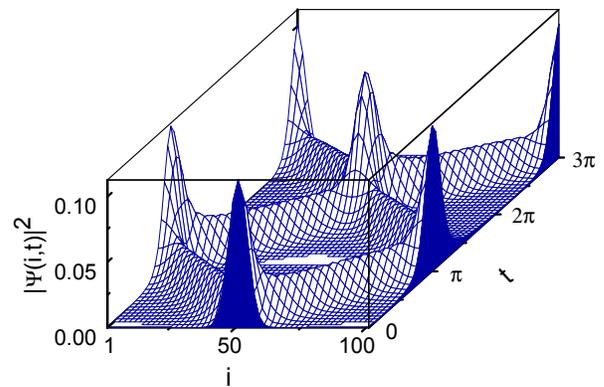}
\caption{\textit{(color online)Time evolution of a Gaussian wave packet of $%
\protect\alpha =0.1$ in a $N=100$ ring system obtained by numerical
simulation. It shows that the two sub-waves are local and keep the shape.}}
\end{figure}

A natural question is that what role the long-range coupling plays for the
peculiar behavior of the propagation of the local wave in such a system. As
shown above, the long-range hopping coupling decays rapidly. Then if the
size of system is large enough, it is believed that the too long range
coupling can be neglected. Thus, the \textquotedblleft long range
coupling\textquotedblright\ in the form (\ref{LR}) can be regarded as the
relative local, or \textquotedblleft microscopic
long-range\textquotedblright\ coupling. In order to investigate this
problem, or the boundary between the so called \textquotedblleft long
range\textquotedblright\ and \textquotedblleft microscopic
long-range\textquotedblright\ couplings,\ we consider the truncated
Hamiltonian

\begin{equation}
H=h_{0}+\sum_{r=odd,\,<r_{0}}J_{r}h_{r},
\end{equation}%
where $r_{0}$\ is the truncation distance. From the above analysis, the
initial state $\left\vert \psi (N_{A},0)\right\rangle $ should recurrence at
the positions $N_{A}$ at instant $2\pi $, if the system is perfect for the
quantum state transfer. The autocorrelation $\left\vert A(t)\right\vert
=\left\vert \left\langle \psi (N_{A},0)\right. \left\vert \psi
(N_{A},t)\right\rangle \right\vert $ is appropriate quantity to investigate
the role that $r_{0}$\ plays. During the period of time $\sim \lbrack 0,3\pi
]$, the maxima of autocorrelations $\left\vert A_{\max }\right\vert =\max
\{\left\vert A(t)\right\vert \}$ of the state (\ref{initial}) in the systems
with $N=500,$ $1000,$ $1500,$ $2000,$ and $r_{0}=10,$ $20,...,100$ are
calculated numerically by the exact diagonalization method. In Fig. 5, the
dependence of the truncation distance $r_{0}$ on the quantity $\left\vert
A_{\max }\right\vert $ is plotted. It shows that $\left\vert A_{\max
}\right\vert $ approaches to $1$ when $r_{0}$ is around 90, which is called
the critical truncation distance or the boundary between the
\textquotedblleft long range\textquotedblright\ and \textquotedblleft
microscopic long-range\textquotedblright\ couplings, for the systems with
different $N$. It indicates that in the case of $N\gg r_{0}$, the wave
packets still travel without spreading. By making use of this observation,
we find that although the interactions between spins are \textquotedblleft
long range\textquotedblright , the $1/r^{2}$-distribution of coupling
strength allow us to limit the maximal interaction range while minimizing
the degradation of the quantum coherence obtained for ideal model. Then we
have the conclusion that the $1/r^{2}$-decay coupling can be regarded as
local coupling, or microscopic long-range\ coupling. In other word, the
robust long-range entanglement between two distant qubits is not due to the
direct long range coupling interaction between them.

From the above observation, we find that, for large size system the
long-range coupling ($r_{0}\gtrsim 90$) can be neglected. Then in
thermodynamic limit, a ring system is equivalent to a more practical system,
open chain system. To demonstrate this, the numerical simulation is employed
to investigate the concurrence $C(l,r_{0},t)$ between two far separated
sites $N/2\pm l$ ($l\sim N/2$) for the $N$-site system with different
truncation $r_{0}$. In Fig. 6, plots of $C(l,r_{0},t)$ for the systems with $%
N=1000$, $l=400$ and different $r_{0}$ are presented. It shows that for an
open chain system, long range entanglement between two distant qubits can be
achieved via the time evolution of a single flip state. The maximal
entanglement created by such system is $0.5$, as measured by the two-point
concurrence.

\bigskip

\begin{figure}[tbp]
\includegraphics[bb=42 281 533 747, width=8 cm, clip]{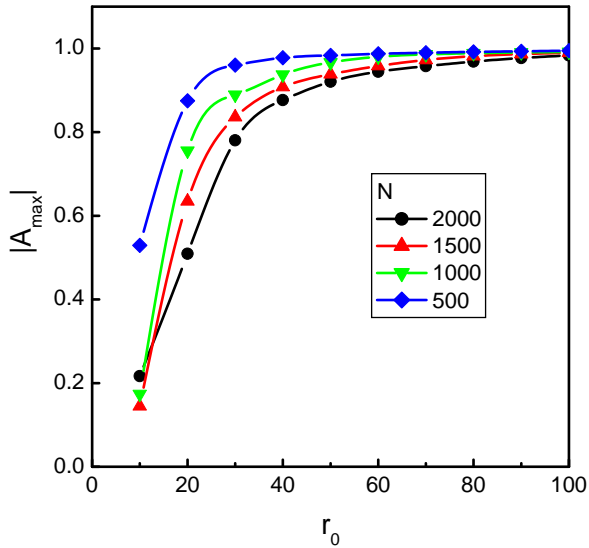}
\caption{\textit{(color online) The maximal autocorrelation functions of the
initial $\protect\delta $-pulse state in the $N$-site systems with the
truncation distance $r_{0}$. It demonstrates that critical $r_{0}$, at which
$\left\vert A_{\max }\right\vert $ start to approach to 1, do not depend on $%
N$ strongly.}}
\end{figure}

\begin{figure}[tbp]
\includegraphics[bb=40 355 490 750, width=8 cm, clip]{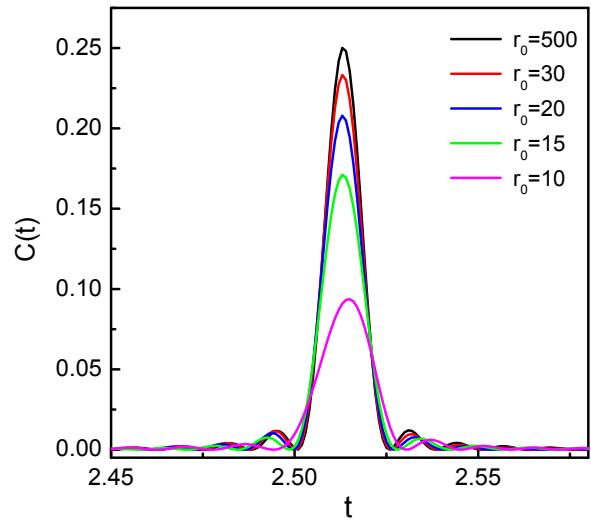}
\caption{\textit{(color online) The time-dependent concurrences of two spins
separated by the distance $2l=800$ in the chain system with $N=1000$, $%
r_{0}=10, 15, 20, 30$ and $500$. The results for $r_{0}=500$ is in
agreement with the analytical analysis. The results for different
truncation approximations show that range of LR coupling can be
taken in a small scale due to the $1/r^{2}$-decay of coupling
constants.}}
\end{figure}

\section{Summary}

In summary, the system with long-range coupling is investigated analytically
and numerically. It is found that the $1/r^{2}$-decay long-range coupling
model can exhibit approximately linear dispersion $\varepsilon \sim
\left\vert k\right\vert $. The dynamics of such model possesses a novel
feather that an initial local wave packet can be separated into two
entangled local wave packets. Furthermore, during the traveling period each
wave packets can keep their shapes without spreading. Numerical simulation
indicates that there exists a critical truncation distance $r_{0}$,\ which
limits the range of the interaction but not affects the generation of
entanglement between two distant qubits in the distance $l\gg r_{0}$. This
model open up the possibility to realize the solid-state based entangler for
creating two entangled but spatially separated qubits.

This work is supported by the NSFC with grant Nos. 90203018, 10474104 and
60433050. It is also funded by the National Fundamental Research Program of
China with Nos. 2001CB309310 and 2005CB724508.

\end{document}